\documentclass[pre,10pt, a4paper,superscriptaddress,nofootinbib,twocolumn,longbibliography,showkeys]{revtex4-1}

\usepackage{float}
\usepackage{bm}
\usepackage[parfill]{parskip} 
\usepackage{graphicx}
\usepackage{amssymb,latexsym} 
\usepackage{amsthm}
\usepackage{amsmath}
\usepackage{color}

\begin{document}

\title{\large Finite Volume \textit{vs.}  Streaming-based  Lattice Boltzmann algorithm for fluid-dynamics simulations:
  a one-to-one accuracy and performance study}  

\author{Kalyan Shrestha}
\email{kalyan.shrestha@polytech-lille.fr}
\author{Gilmar Mompean}
\author{Enrico Calzavarini}
\affiliation{Laboratoire de M\'ecanique de Lille CNRS/UMR 8107, Universit\'e Lille 1 \& Polytech'Lille,\\
Cite Scientifique, Av. P. Langevin, F-59650 Villeneuve d'Ascq, France}

\date{\today}

\begin{abstract}
A new finite volume (FV) discretisation method for the Lattice Boltzmann (LB) equation which combines high accuracy with limited computational cost is presented.
In order to assess the performance of the FV method we carry out a systematic comparison, focused on accuracy and  computational performances, with the standard \textit{streaming} (ST)  Lattice Boltzmann equation algorithm. To our knowledge such a systematic comparison has never been previously reported. In particular we aim at clarifying whether and in which conditions the proposed  algorithm, and more generally any FV algorithm, can be taken as the method of choice in fluid-dynamics LB simulations. For this reason the comparative analysis is further extended to the case of realistic flows, in particular thermally driven flows in turbulent conditions. We report the first successful simulation  of high-Rayleigh number convective flow performed by a Lattice Boltzmann FV based algorithm with wall grid refinement.    
\end{abstract}


\maketitle

\section{Introduction\label{introduction}}
Since its introduction, the Lattice Boltzmann (LB) equation method for fluid-dynamics simulations has enjoyed increasing success \cite{Succi2001book}.
Reasons are two-fold, on one hand the mesoscopic level of description on which it is based  goes beyond the Navier-Stokes continuum matter description of fluids and it eases, as compared to other macroscopic methods, to accommodate for complex effects such  as for instance the interaction between different fluid components, phase-change processes, non-newtonian rheology. The extensions of the LB methods in such directions are countless (\textit{e.g.} multiphase and multicomponent flows \cite{Gunstensen, Shan, Flekkoy1, Flekkoy2, Swift, Chen}; flows with  suspensions \cite{Ladd1, Ladd2, Ladd3, Xia, Connington}; emulsions \cite{Boghosian}; porous media \cite{Gunstensen2, Kang, Zhang2, Guo2}; 
  natural convection \cite{ShanRB}; reactive transport \cite{Kang1, Kang2}; combustion \cite{Chen10, Yamamoto}; magneto-hydrodynamics \cite{Dellar2002, Chen5}). On the other hand, the LB method has also very appealing features from a computational point of view. It is simple to implement, free of numerical diffusion and stability issues 
 and suitable for parallelization due to its local-in-space character.\\
However, when it comes to the simulation of turbulent flows one shortcoming of the method, \textit{the limitation to equispaced grids}, becomes evident. We recall here that a developed turbulent flow in the presence of any sort of bounding geometry (or any local forcing term) develops space inhomogeneities and as such grid refinement in numerics becomes necessary. It shall be made clear that in such a context, grid-refinement is not an additional requirement in order to increase the accuracy of a simulation but is rather an unavoidable need in order to save memory usage and computational power and being able to access higher - read more realistic - turbulent flows regimes.  In summary any state-of-the-art computational fluid dynamics method (CFD) calls for grid refinement.\\ 
Several approaches have been proposed  in order to overcome the shortcoming of equi-spaced grids in Lattice Boltzmann (LB) equation simulations.
Here we mention, i) the grid refinement methods which make use of locally nested equispaced grids \cite{Filippova}, ii) the techniques based on off-lattice interpolation schemes \cite{he-jcp-1996,he-jcp-1997} iii) the finite-difference \cite{GuoPRE2003}, finite-volume \cite{Benzi-rep-1992,Nannelli} or finite-elements \cite{LeeJCP2001} LB discretisation methods, and  iv)  the extension of the LB equation to general manifolds \cite{Mendoza2013}. 
There are however important drawbacks, all such reformulation are computationally more expensive, or introduce extra stability limitations enforced by space/time discretisation which were not present in the original so-called \textit{streaming based} implementation.  Presently, the only viable way  seems to be the nested monospaced grid method (i), which has allowed to simulate  turbulent channel flows \cite{Touil} and even more complex flow geometries \cite{geller2013}. However, in this case the advantage in terms of accuracy and efficiency compared to state of the art direct numerical simulation (DNS) \textit{e.g.} spectral methods is limited.

This paper presents a new Finite-Volume (FV) discretisation method for the Lattice Boltzmann equation which besides a high level of accuracy also displays a contained computational cost.  In order to assess the performance of this new FV method we carry out a systematic comparison with the standard streaming (ST) formulation. To our knowledge such a methodical comparison of accuracy and computational performances has not been made previously. We aim at clarifying whether and in which conditions the proposed FV algorithm can be taken as the method of choice in fluid-dynamics LB simulations.\\
The paper is organised as follow. In the next section we describe two different discretizations of the LB equation. To begin with, we briefly review the key points of the streaming implementation (\S \ref{sec:methodST}), then we present the new FV based formulation (\S \ref{sec:methodFV}). Our guidelines in the  development of the new FV method are the the simplicity and computational efficiency of the implementation, yet retaining a level of accuracy which takes the  ST method as the baseline. 
Results of this study are reported in \S \ref{sec:acc}. First, we address the accuracy comparison, later on the computational efficiency of the algorithm. A further section (\S \ref{sec:rb}) takes the analysis to more realistic flows, in particular thermal flows in turbulent conditions. To our knowledge, for the first time a high-Rayleigh number convective flow with wall grid refinement is simulated by a LB based code. Final remarks and perspectives are reported in the conclusions.

\section {Method}\label{sec:method}

We focus here on the LB method  with  the Bhatnagar-Gross-Krook (BGK) collision operator, which is characterised by a single relaxation time $\tau$ towards a local equilibrium state. The equation of motion reads: 
\begin{equation}\label{eq:LB}
\frac{\partial f_{\alpha}}{\partial t}+ {\bm c}_{\alpha} \cdot {\bm \nabla} f_{\alpha} = \frac{1}{\tau} (f_{\alpha}^{eq} - f_{\alpha}) + F_{\alpha} \quad  \alpha = 0,\ldots , N_{pop}
\end{equation}
where $f_{\alpha}({\bm x},t) $  is one of the $N_{pop}$ distribution functions for particles (also called \textit{populations})
with velocity ${\bm c}_{\alpha}$ at position $\bm{x}$ and time $t$. 
The set of $f_{\alpha}$ distribution functions relax towards a local equilibrium state $f_{\alpha}^{eq}({\bm x},t)$ which is prescribed in terms of local macroscopic variables (in non thermal models, as here, they are just the fluid velocity and density) \cite{Succi2001book}. The macroscopic fluid mass and momentum density can be computed as $\rho = \Sigma_{\alpha} f_{\alpha}$  and $\rho \bm{u}= \Sigma_{\alpha} \bm{c}_{\alpha} f_{\alpha}$ and  the kinematic viscosity as $\nu = \tau\ c_s^2 $, where the constant $c_s$  stands for the so-called lattice speed of sound, whose value depends on the specific velocity lattice topology (see for instance Succi's book \cite{Succi2001book} for details).
Finally, $F_{\alpha}$ is a forcing term, constructed in such a way to model the effect of a macroscopic body force term.\\ 
Note that eq. (\ref{eq:LB}) is discrete just in the velocity space ( for this reason it is also known as discrete velocity Boltzmann equation). Up to this level no discretisation has been taken neither in the spatial domain or in the temporal one.  Such further discretisations can take different paths as we describe in the following sections.

\subsection {Outline of the Streaming Lattice Boltzmann algorithm}\label{sec:methodST}
It is here useful to briefly recall the steps that have to be made in order to obtain from eq.(\ref{eq:LB}) the standard streaming (ST) LB algorithm.
First, by applying to the above partial differential equation (PDE) the technique of the characteristics  along the lines $\bm{x}(t) =\bm{x}(0) + \bm{c}_{\alpha}\ t\ $, one gets the ordinary differential equation (ODE)
\begin{equation}\label{eq:ode}
 \frac{d}{dt} f_{\alpha} = \frac{1}{\tau} (f_{\alpha}^{eq} - f_{\alpha}) + F_{\alpha}.
\end{equation}
Second, the discrete integration in time, of step $\Delta t$, is performed by applying the semi-implicit Crank-Nicolson method.
Such a step is followed by a convenient redefinition of the distribution functions for the lattice populations $\tilde{f}_{\alpha} =f_{\alpha} - \frac{\Delta t}{2\tau} ( f_{\alpha}^{eq} - f_{\alpha} + \tau F_{\alpha})$ which makes the scheme explicit, leading to \cite{he-jcp-1998}:
\begin{widetext}
\begin{equation}\label{eq:ST}
\tilde{f}_{\alpha}(\bm{x}+\bm{c}_\alpha \Delta t , t + \Delta t) = \tilde{f}_{\alpha}(\bm{x}, t)+ \frac{\Delta t}{\tilde{\tau}} (\tilde{f}_{\alpha}^{eq}(\bm{x}, t) - \tilde{f}_{\alpha}(\bm{x}, t)) + \Delta t  \left( 1 - \frac{\Delta t}{2 \tilde{\tau}} \right) F_{\alpha}
\end{equation}
\end{widetext}
where $\tilde{\tau} = \tau + \Delta t / 2$ is a redefined relaxation time $(\tilde{\tau} >\Delta t / 2)$. 
It is easy to derive the relations between the macroscopic variables and the tilded ($\sim$) quantities, they are respectively:   $\rho = \Sigma_{\alpha} \tilde{f}_{\alpha}$, $\rho \bm{u}= \Sigma_{\alpha} \bm{c}_{\alpha} (\tilde{f}_{\alpha} + \frac{\Delta t}{2} F_{\alpha})$ and $\nu = (\tilde{\tau} - \Delta t /2) c_s^2 $.
Note that a factor $ 1 - \frac{\Delta t}{2 \tilde{\tau}} $ in front of the forcing term needs to be introduced \textit{a posteriori} in order for the discretised eq.(\ref{eq:ST}) to give the same hydrodynamics limits as (\ref{eq:LB}) \cite{Guo}.\\
 The numerical implementation of (\ref{eq:ST}) is straightforward. It can be divided in two steps: i) the computation of the right-hand-side and  ii) the displacement (or streaming) of the computed values on the lattice according to the direction and intensity of $\bm{c}_{\alpha}$. It is important to note that the integration along the characteristics introduces a link between the space and time discretization,  which reads  $\bm{c}_{\alpha} = \Delta \bm{x}_{\alpha} / \Delta t$.
This means that if one choose the cartesian components of the set of $\bm{c}_{\alpha}$ velocities to be either $\pm 1$ or $0$,  it implies that $\Delta x_{\alpha,i} = \Delta t $ or $0$. The standard choice (but not the only possible one) is $ \Delta x_{\alpha,i} = \Delta t = 1$ \cite{Succi2001book}.

\subsection {Lattice Boltzmann Finite Volume Formulation}\label{sec:methodFV}
The method of characteristics is very convenient from a computational point of view because it reduces the complexity of the integration of a PDE to a simple ODE, however at the same time it introduces \textit{a tight link between the shape of the velocity lattice and the spatial discretisation mesh}. Such a constraint can be removed if one takes the more usual numerical approach based on i) a direct spatial discretisation of eq. (\ref{eq:LB}) combined with ii) an independent time discretisation phase. For the first step, several standard  options are  available, such as finite elements, finite differences or finite volumes methods. \\
 The idea of using a finite volume method to decouple  the spatial numerical mesh from the velocity lattice structure  was first proposed by Nannelli and S. Succi, \cite{Nannelli} (see also \cite{Benzi-rep-1992}), in this seminal paper a low-order upwind scheme was suggested for the  discretisation of the advection (or flux) term.  The idea was further refined in Amati \textit{et al.} \cite{Amati}, where piece-wise linear interpolation scheme was suggested for the treatment of the flux term.  While these first works were limited to stretched cartesian grids, Chen \cite{Chen20} presented a volumetric formulation, based on a cell-centered discretisation scheme, which allowed for the adoption of arbitrary \textit{structured} meshes. The formulation was further developed by Peng \textit{et al.} \cite{Peng1,Peng2} through cell-vertex FV scheme, which displayed enhanced stability properties. Sbragaglia \& Sugiyama  \cite{Sbragaglia2010} applied Peng's scheme to an energy-conserving LB model to study for the first time thermal convective flows. More recently Ubertini \textit{et al.} \cite{Ubertini} addressed the problem of \textit{unstructured} bidimensional triangular meshes, which allow great flexibility on one hand,  but  also reintroduce known issues related to numerical stability. This has been further refined in a work by Zarghami \textit{et al.} \cite{Zhargami} through a cell-centered FV approach on arbitrary mesh in two dimension.  
 A Total Variation Diminishing (TVD) formulations for LB FV algorithm has been suggested by Patil et al. \cite{Patil}, where stability and accuracy can be efficiently enhanced.
 Despite all these contributions, at present the situation is still far from being solved. If on one side it has been shown that a satisfactory level of precision can be reached by the FV method on the other hand this is often at the price of the high computational costs needed to obtain a stable algorithm. As an example in a recent work \cite{Zhargami}, where a series of laminar but relatively complex flows  over non-homogeneous meshes were simulated, a fifth-order Runge-Kutta scheme had to be adopted for the time discretisation in order to have stable results. The consequence on  the computational cost is evident since in such a scheme the advection terms of (\ref{eq:LB})  need to be computed five times per time step (while in the ST method it is performed just by means of a memory shift, the streaming). As a consequence the FV LB is rarely a method of choice in fluid-dynamics simulations (see also the discussions in \cite{Stiebler} and in \cite{Pietro}).  

The present paper further develops the finite volume Lattice Boltzmann method in order to simulate fluid flow problems with higher accuracy, greater stability properties and comparable performance as the ST method. The FV method that  we propose is of the type denoted as cell-centered (as opposed to vertex-centered, see Fig. \ref{fig:FV}a). Its most original features concern the approach taken for the time discretisation (\S \ref{sec:time}) and the method of  fluxes computation which adopt, for the first time in this context, a quadratic upwind scheme (\S \ref{sec:adv}). In the following we detail the steps taken in developing it.\\

\subsubsection {Space discretisation}\label{sec:space}
Upon integration of eq. (\ref{eq:LB}) over a volume $V$ (of surface  $S$) and by applying the flux theorem we get 
\begin{widetext}
\begin{equation}
 \int_V \frac{\partial f_{\alpha}}{\partial t} \ dV + \int_S {\bm c}_{\alpha} \cdot {\bm n} \ f_{\alpha}\ dS = \int_V \frac{1}{\tau} (f_{\alpha}^{eq} - f_{\alpha})\ dV +   \int_V F_{\alpha} \ dV
\end{equation}
\end{widetext}
We then assume that every term in the volume integrals can be considered as constant and its magnitude taken at a reference location ${\bf x}$ (also called \textit{node}) inside $V$.\\ The term in the surface integral however, carries some kind of spatial variability.  When such a surface is decomposed in $M$ faces (as in a structured grid of nodes with connectivity index $M$) it is convenient to make the assumption that $f_{\alpha}$ is constant on each of the $S_j$ surfaces perpendicular to ${\bm n}_j$ and denoting its value with $\left[ f_{\alpha} \right]_j$. This altogether leads to:
\begin{equation}\label{eq:fv}
 \frac{\partial f_{\alpha}}{\partial t}   +  \frac{S_j}{V} {\bm c}_{\alpha} \cdot  {\bm{n} }_j  \left[ f_{\alpha} \right]_j = \frac{1}{\tau} ( f_{\alpha}^{eq} - f_{\alpha}) + F_{\alpha}
\end{equation}
Where summation over the repeated index $j$ is applied. 

\subsubsection {Time discretisation}\label{sec:time}
If the time derivative is discretized by the explicit Euler scheme, we get: 
\begin{widetext}
\begin{equation}\label{eq:euler}
 f_{\alpha}^{(t+\Delta t)}  = f_{\alpha}^{(t)}- \Delta t  \frac{S_j}{V} {\bm c}_{\alpha} \cdot  {\bm{n} }_j  \left[ f_{\alpha}^{(t)} \right]_j + \frac{\Delta t }{\tau} ( f_{\alpha}^{eq\, (t)} - f_{\alpha}^{(t)}) + \Delta t \ F_{\alpha}, 
\end{equation}
\end{widetext}
where the superscript indexes $(t)$ and $(t+\Delta t)$ denotes respectively the current and the next discrete time instant.  Such an approach however, puts tight bounds on the maximum allowed $\Delta t$. It is easy to show that if we discard  the advection and the forcing terms and assume $f^{eq}$ to be constant, the stability region of the method is $0 < \Delta t \leq 2 \tau$. Empirically it is possible to show that this range becomes even narrower when the non-local advection term, the forcing and the time dependency in $f^{eq}$ are taken into account. 
The fact that $\Delta t_{max}$  depends  on and is bounded by the value of $\tau$ is  a known problem in FV LB implementations. It poses, among others,  a severe limitation for the simulations of turbulent flows (\textit{i.e.} low viscosity flows). On the opposite, such a constraint does not exist in the ST approach (where $\Delta t$ is independent of $\tau$).  
Different solutions have been proposed in the literature, often resorting explicit time discretisation schemes of higher order, for example multi-stages Runge-Kutta schemes.  However, as we mentioned above such schemes only produce marginal improvements at the expenses of considerably increasing the computations. The Runge-Kutta schemes for example requires multiple evaluations of the full right-hand-side terms on (\ref{eq:fv}). 
We opt for a different approach, with a better trade-off  between the enhancement of the stability limit for $\Delta t$ and the growth in computational cost. 

Similarly to what is done for the classic LB streaming method, in the steps from eq.(\ref{eq:ode}) to (\ref{eq:ST}), a possible improvement consists in taking also for the FV algorithm a semi-implicit integration scheme.  However, this is not directly possible for eq.(\ref{eq:fv}) because of the presence of the advection term. Therefore,  we propose to limit such an approach only to the collision and forcing terms, after few manipulations (more details in the appendix) one gets the discretised form:
\begin{widetext}
\begin{equation}\label{eq:semi} 
\tilde{f}_{\alpha}^{(t+\Delta t)} = \tilde{f}_{\alpha} - \Delta t  \frac{S_j}{V} {\bm c}_{\alpha} \cdot  {\bm{n} }_j  \left[ \tilde{f}_{\alpha} +\frac{\Delta t}{2 \tilde{\tau}}( \tilde{f}_{\alpha}^{eq} - \tilde{f}_{\alpha} ) + \frac{\Delta t}{2} F_{\alpha} \right]_j + \frac{\Delta t}{\tilde{\tau}} (\tilde{f}_{\alpha}^{eq} - \tilde{f}_{\alpha}) + \Delta t \left( 1 - \frac{\Delta t}{2 \tilde{\tau}} \right)  F_{\alpha}
\end{equation}
\end{widetext}
The above equation share the same definitions of  (\ref{eq:ST}) for the tilded distribution function, $\tilde{f}_{\alpha}$  and the relaxation time ($\tilde{\tau}$).
The rule of computing the macroscopic fields  and the viscosity  $\nu = \tau\ c_s^2  = \left( \tilde{\tau} - \Delta t / 2 \right) c_s^2$ are exactly the same as for the ST algorithm. Correspondingly, the term $ 1 - \frac{\Delta t}{2 \tilde{\tau}} $ in front of the forcing has been introduced \textit{a posteriori} to keep the same hydrodynamic limit.
However, one can immediately note the advected field in the equation is not simply a distribution function but a rather a complex term involving also the equilibrium distribution and the forcing. 
The main advantage of this approach is that a stability analysis under the same hypothesis mentioned above (neglecting advection, forcing and time dependences in the equilibrium function) shows now that  every time step length $\Delta t$ is stable. However, we find that the situation reached so far is not yet satisfactory. From simple numerical tests we observe that even with this discretisation scheme the time-step size  is still restricted by the relaxation time, particularly for small relaxation time values. The origin of this still limited stability of the scheme lies now in the advection term.  
For this reason a further  refinement is proposed.  We set it into place by applying the so-called Heun predictor-corrector scheme to  the advection term. In other words we use the calculation of the population based on  (\ref{eq:semi}), now called $\tilde{f}^{*}$, as an intermediate value for  constructing an explicit trapezoidal integration rule applied to the advection:

\begin{widetext}
\begin{eqnarray}\label{eq:semi-heun}\nonumber
 \tilde{f}_{\alpha}^{(t+\Delta t)} \mkern-10mu \ & = &\  \mkern-10mu \ \tilde{f}_{\alpha}   - \Delta t \frac{S_j}{V} {\bm c}_{\alpha} \cdot  {\bm{n} }_j   \frac{\left[ \tilde{f}^{*}_{\alpha} +\frac{\Delta t}{2 \tilde{\tau}}( \tilde{f}^{eq *}_{\alpha} - \tilde{f}^{*}_{\alpha} ) + \frac{\Delta t}{2} F^{*}_{\alpha} \right]_j   + \left[ \tilde{f}_{\alpha} +\frac{\Delta t}{2 \tilde{\tau}}( \tilde{f}_{\alpha}^{eq} - \tilde{f}_{\alpha} ) + \frac{\Delta t}{2} F_{\alpha} \right]_j}{2} \\
 & +&   \frac{\Delta t}{\tilde{\tau}} (\tilde{f}_{\alpha}^{eq} - \tilde{f}_{\alpha}) + \Delta t \left( 1 - \frac{\Delta t}{2 \tilde{\tau}} \right)  F_{\alpha}.
\end{eqnarray}
\end{widetext}

where $F^{*}_{\alpha}$ indicates the LB forcing term computed from $\tilde{f}^{*}$.
This scheme enjoys greater stability at the additional computational price of a  second evaluation of the advection term.  In order to make this observations more quantitative we should  first specify the way in which the flux terms $\left[ \ldots \right]_j$  are computed. Indeed, the exact stability properties of the method depends upon the implementation of the advection term, 
that we discuss in the following section. 
\begin{figure}[hbt]
  \centering
              \includegraphics[width=0.45\textwidth]{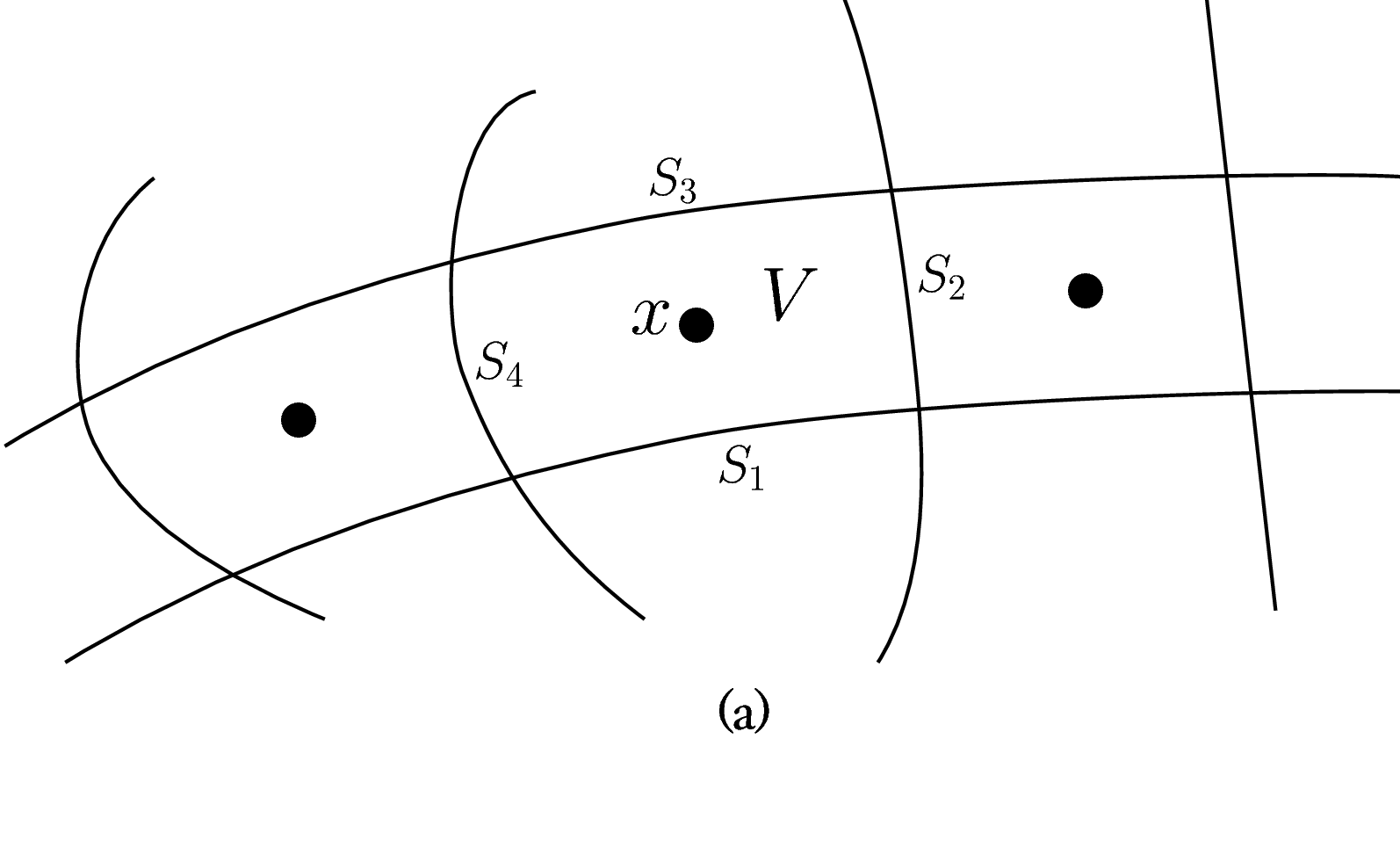}
              \includegraphics[width=0.47\textwidth]{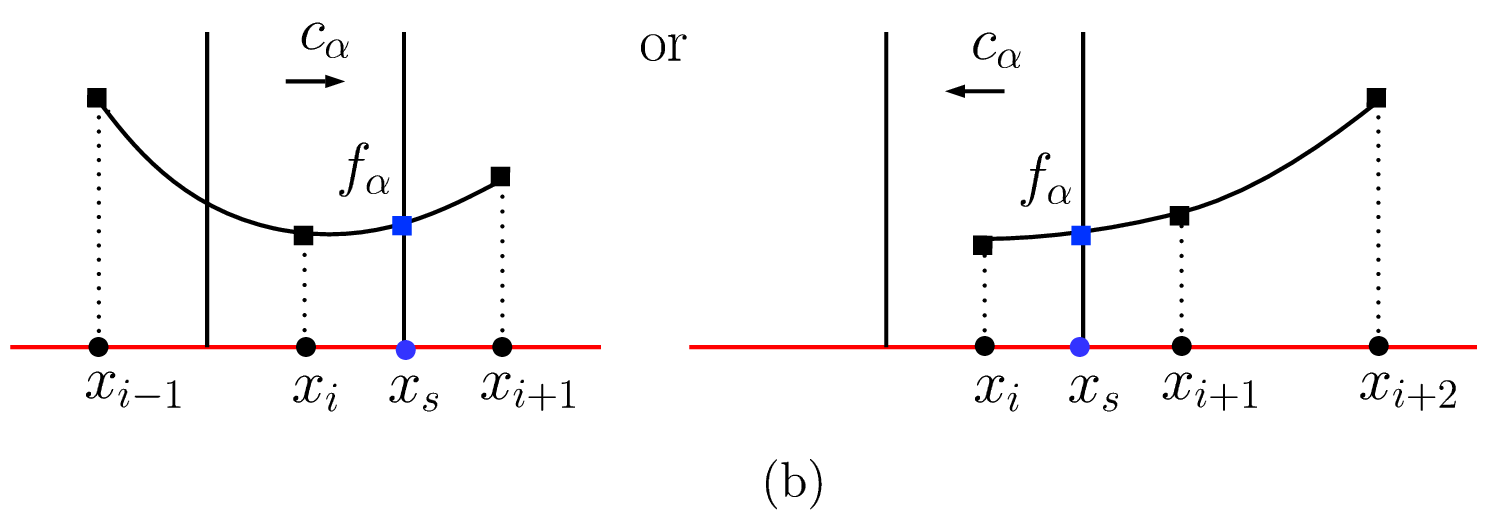}
  \caption{(a) Cartoon of the finite volume space discretisation: the dot denotes the position of the cell center (where the value of $f_{\alpha}(x)$ is defined), while the lines marks the cell boundaries. The cell has volume $V$ and each boundary surface is denoted with $S_j$ with $j=0,\ldots, 3$ in two dimensional space. (b) Sketch of the quadratic upwind interpolation scheme (QUICK) for the estimating the value of $f_{\alpha}$ at the cell boundary position $x_S$. Note that the interpolation method make use of different nodes according to the direction of the population velocity $\vec{c}_{\alpha}$.}
  \label{fig:FV}
\end{figure}

\subsubsection {Approximation of the advection term}\label{sec:adv}
There exist several ways to estimate the non-local term $ \left[ f_{\alpha} \right]_j$ and each one can be characterised by a spatial order of accuracy.   The complexity of such an estimation also depends on the grid characteristics. 
Even for structured but irregular grids an high-order estimation of $ \left[ f_{\alpha} \right]_j$ becomes expensive in computation terms. In order to simplify such a problem, we limit the following discussion to the case of structured regular grids, that is to say to the case where the nodes lie on lines. This is the case for instance of a non-uniform cartesian grid (the typical case of wall-refinement), but it also apply to a uniformly skewed non-orthogonal grids.

It has been long known that fluxes in advection equations are better approximated by upwind schemes, which are interpolation schemes biased in the direction determined by the sign of the characteristic speeds (the set of ${\bf c}_{\alpha}$ in our case).  At the lowest order of accuracy, and easiest level of implementation, there exist the first order up-wind scheme, 
increasing the refinement leads to linear interpolation schemes or even to more refined quadratic schemes (which are of $3^{rd}$ order of spatial accuracy).
While low-order schemes introduces artificial numerical dissipations, higher-order ones lead to spurious oscillations, especially evident near the boundaries.  
This is also true in the present cell-centered FV implementation, in particular  zero-order or linear up-wind interpolation schemes leads to inaccurate results. Even a cell-centered symmetric schemes, which here does not display extra dissipation, produces inaccurate results in the presence of boundaries. Empirically, we find that the quadratic upstream interpolation, known as QUICK method \cite{Leonard1979}, is the simplest one to give accurate results both in open (\textit{i.e.} periodic) and bounded domains.

According to this approach, on each surface $S_j$ at position say ${\bm x_{S_j}}$, $ \left[ f_{\alpha} \right]_j$ is approximated  via a combination of the value of $f_{\alpha}$ in the two nodes bracketing the surface  (denoted with ${\bm x}$ and ${\bm x^+}$) and a third node that is located upstream respect to  direction of the projection of  $\hat{{\bm c}_{\alpha}}$ on  $\hat{\bm{n} }_j$ (denoted either ${\bm x^{++}}$ or ${\bm x^{-}}$). The interpolant function is a parabola $a+b\ \xi + c \ \xi^2$, with $\xi$ the linear coordinate spanning on the line connecting the nodes (see sketch in Fig. \ref{fig:FV}b).  This leads to:
\begin{widetext}
\begin{eqnarray}\nonumber
  \left[ f_{\alpha} ({\bm x_{S_j}}) \right]_j &=& (1-\gamma_1+\gamma_2)f_{\alpha}({\bm x})  +  \gamma_1 \ f_{\alpha}({\bm x^{+}})  -  \gamma_2\ f_{\alpha}({\bm x^{-}}) \mid_{\alpha:\ \hat{{\bm c}_{\alpha}} \cdot \hat{\bm{n} }_j  > 0}\\
&+&  (1-\gamma_3+\gamma_4) f_{\alpha}({\bm x^{+}})  + \gamma_3\ f_{\alpha}({\bm x}) -  \gamma_4 \ f_{i}({\bm x^{++}}) \mid_{\alpha :\ \hat{{\bm c}_{\alpha}} \cdot \hat{\bm{n} }_j  < 0}
\end{eqnarray}
\end{widetext}
where the  $\gamma_{1,2,3, 4}$ are 4 coefficients, that shall be evaluated and/or stored for each surface of the control volumes.

\subsubsection {Force term}
Finally a brief remark on the forcing term $F_{\alpha}$ in the LB equation. The simplest way to implement it, is by the expression:
\begin{equation}\label{eq:F1}
F_{\alpha} =    w_\alpha \ \frac{{\bm c}_{\alpha} \cdot \rho\ {\bm a}}{c_s^2}
\end{equation}
where the summation over index $\alpha$ is not implied, $w_\alpha$ is a lattice dependent weight and the product $\rho {\bm a}$ represents  the force per unit volume in physical space (for example in case of a gravitational external field ${\bm a} = {\bm g}$). 
The above expression satisfies the conditions $\Sigma_{\alpha} F_{\alpha} = 0$ and $\Sigma_{\alpha} {\bm c}_{\alpha} F_{\alpha} = \rho\ {\bm a}$, which are required for  eq. (\ref{eq:LB}) to give the correct macroscopic effect of a body force term.
However, when the body force is time/space dependent and  eq. (\ref{eq:LB}) is discretised in space and time, such as in (\ref{eq:ST}), the above expression needs to be refined in order to remove spurious discretisation terms that would otherwise appear in the macroscopic limit.
The corrected expression, first proposed by Guo \textit{et al.} \cite{Guo}, is
\begin{equation}\label{eq:F2}
F_{\alpha} =    w_\alpha \left(  \frac{{\bm c}_{\alpha} - \bm{u} }{c_s^2}   + \frac{(\bm{c}_{\alpha} \cdot \bm{u})\ \bm{c}_{\alpha}}{c_s^4} \right)   \rho\ {\bm a}.
\end{equation}
with an overall multiplicative factor $1-\Delta t/(2\tau)$ when the distribution functions $\tilde{f}_{\alpha}$ are evolved in place of $f_{\alpha}$.
Note that accordingly (by employing the relation between $\tilde{f}_{\alpha}$ and  $ \rho \bm{u}$ and $\rho$ given in \S \ref{sec:methodST}) one gets the fluid velocity as $ \bm{u}= \Sigma_{\alpha} \bm{c}_{\alpha} \tilde{f}_{\alpha} / \Sigma_{\alpha} \tilde{f}_{\alpha}+ \frac{\Delta t}{2} {\bm a}$.

\subsubsection {Boundary conditions}
In the following we consider the implementation two types of  boundary conditions (BC): i) no-slip walls and ii) fixed density (or equivalently pressure) boundaries. 
The physical domain boundaries lie on the faces of the external control volumes. Similarly to the bounce-back approach for the streaming LB algorithm, we introduce  in-wall ghost cells. However, in the QUICK treatment of the advection two ghost cells are needed instead of one. The ghosts cells are located in-wall and have centres at position mirroring the first and second nodes in the fluid domain. 
 Let's suppose  that the quantity to be advected  is $f_{\alpha}$ and that the boundary condition is to be imposed on the $S$ cell surface, whose center is at ${\bm x}_S$.
 For simplicity we assume that $S$ lies along the plane $(y,z)$ perpendicular to $x$, with the $x$ axis pointing inward (\textit{i.e.} in the fluid bulk direction).  Consequently, the first two nodes in the fluid domain are located at position  ${\bm x}_1 = {\bm x}_S + \Delta x_1/2$  and ${\bm x}_2 = {\bm x}_S + \Delta x_1 + \Delta x_2/2$, where $\Delta x_1$ and $\Delta x_2$ represents the linear size of the two first discretisation volumes. Accordingly, the ghosts cells are at positions ${\bm x}_{-1} = {\bm x}_S - \Delta x_1/2$ and ${\bm x}_{-2} = {\bm x}_S - \Delta x_1 - \Delta x_2/2$.
 A no-slip boundary condition  requires ${\bm u}({\bm x}_S) = 0$, while the density at $\rho({\bm x}_S)$ is free to take any arbitrary value.  This corresponds to the constraint $\Sigma_{\alpha} {\bm c}_{\alpha} f_{\alpha}({\bm x}_s) = 0$.  The simplest way (but not the only one) to enforce it, is to set the in-wall nodes as the following:
 \begin{eqnarray}
f_{\alpha}({\bm x}_{-1}) &=& f_{inv(\alpha)}({\bm x}_{1}),  \nonumber \\
\quad f_{\alpha}({\bm x}_{-2}) &=& f_{inv(\alpha)}({\bm x}_2 ) ,
\end{eqnarray}
where $inv(\alpha)$ is an integer valued function that selects the population moving along the opposite direction with respect to ${\bm c}_{\alpha}$. We have verified that such a choice does not introduce artificial fluctuations at the boundary, that would quickly generates instabilities.  This implementation of BC, that we dub \textit{double reflection}, has a  first-order of accuracy in space (it does not implement the quadratic interpolation) and therefore it leaves room for further improvements.

If instead we are interested to impose a density value at the border, say $\rho_S = \Sigma_{\alpha}  f_{\alpha}({\bm x}_S)$, we need to resort an extrapolation strategy. We proceed as follow, first the density value  $\rho({\bm x}_{-1})$ is linearly extrapolated from the values $\rho_S$ and $\rho({\bm x}_1)$, similarly $\rho({\bm x}_{-2} )$ is derived from $\rho_S$ and $\rho({\bm x}_{2})$. Second, we assign the in-wall the distribution functions as follows  
 \begin{eqnarray}
f_{\alpha}({\bm x}_{-1}) &=& \frac{\rho({\bm x}_{-1})}{\rho({\bm x}_1)}f_{\alpha}({\bm x}_{1}),  \nonumber \\ 
f_{\alpha}({\bm x}_{-2}) &=& \frac{\rho({\bm x}_{-2})}{\rho({\bm x}_2)}f_{\alpha}({\bm x}_{2}). 
\end{eqnarray}
Also the above choice, a rescaling of the bulk distribution functions, is not the only viable way for the implementation of fixed density BC, however it is one that has revealed to not to introduce wall disturbances. 
As a final remark, we shall note that in the implementation of (\ref{eq:semi-heun}) the boundary conditions  need not to be implemented on the redefined distribution function $\tilde{f}_{\alpha}$ but rather on the original $f_{\alpha} = \tilde{f}_{\alpha} +\frac{\Delta t}{2 \tilde{\tau}}( \tilde{f}_{\alpha}^{eq} - \tilde{f}_{\alpha} ) + \frac{\Delta t}{2} F_{\alpha}$.  

The algorithm presentation given so far is independent of the particular  microscopic velocity lattice topology. In the present work and for the accuracy study presented in the reminder of this manuscript we make the choice to always use the so called $D3Q19$ lattice, which is a standard option for three-dimensional LB simulations and reduces to the $D2Q9$ lattice for two-dimensional flow problems \cite{Succi2001book}.

\begin{figure}[hbt]
  \centering
              \includegraphics[width=0.5\textwidth]{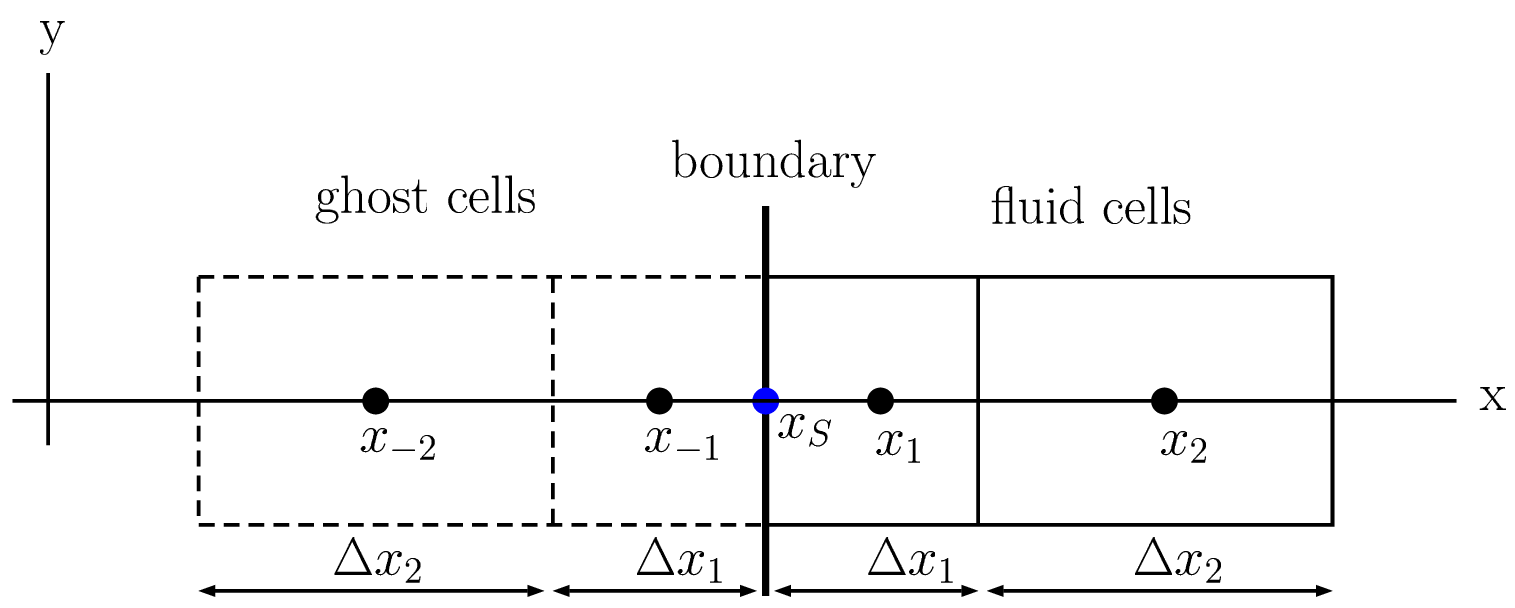}
  \caption{Illustration of the finite-volume arrangement for the implementation of the \textit{double-reflection} boundary conditions.}
  \label{fig:BC}
\end{figure}

\section {Accuracy tests}\label{sec:acc}
In this section we address the accuracy of the present LB FV algorithm and we compare it with the ST algorithm. In particular we approach the following questions: i)  to which degree the FV algorithm correctly describes the dynamics of a low Reynolds number viscous flow? Which is its order of spatial accuracy  and how does it compare with ST? 
ii)  Is there any optimal usage of the FV algorithm in order to take advantage of the grid refinement and obtaining highly accurate solutions?

\subsection {Viscosity evaluation}
\begin{figure}[hbt]
  \centering
             \includegraphics[width=0.5\textwidth]{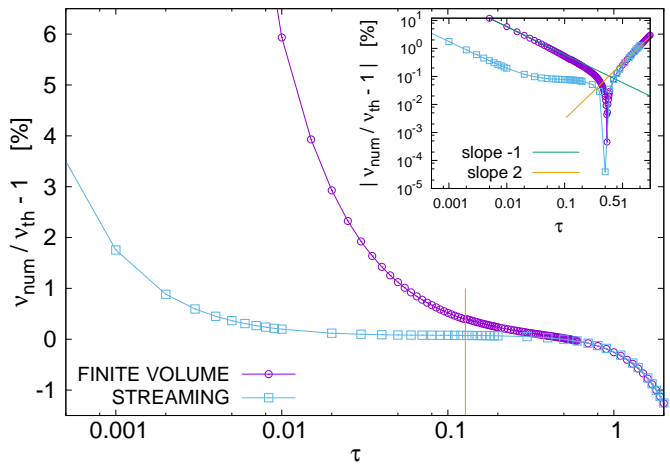}
  \caption{Relative error of measured kinematic viscosity $\nu_{\textrm{num}}$ respect to the expected one $\nu_{th} = \tau\ c_s^2$ as a function of the relaxation time $\tau$.  In the finite volume case $\Delta t = 1$ for $\tau \geq 0.13$ (marked with a vertical line) and $\Delta t = 0.1$ for $\tau < 0.13$, while in the Streaming case $\Delta t = 1$ always. In the inset, the absolute value of the same error in log-log scale.}
  \label{fig:KOL}
\end{figure}

A simulation is performed on a physical domain of size, $[L_x,L_y,L_z]=[1,64,1]$. For this test, the number of grid nodes per direction (indicated with $N_{x,y,z}$)  is also the same. The flow is initialised with an one-dimensional sinusoidal velocity amplitude profile of the form 
\begin{equation}
{\bf u}(x,y,z) = \left( u_x(y) , u_y ,u_z \right) = \left( A \ \sin \left(\frac{2 \pi\ y}{ L_y} \right) , 0 , 0  \right)
\end{equation}
and it is left to decay in time. We monitor the behaviour of the total kinetic energy  in time, $k_{tot}(t)$, which is expected to decrease exponentially as $k_{tot}(t) = \frac{1}{4} A^2 L_y^2 \ e^{-2 (2 \pi/L_y)^2 \nu \ t }$, with  $\nu$ representing the fluid kinematic viscosity.  The reproduced value of $\nu$ can be deduced from a least-square fit of $k_{tot}(t)$ and then compared to the theoretically expected value  $\nu = \tau\ c_s^2   = \left( \tilde{\tau} - \Delta t / 2 \right) c_s^2$.  
The degree of accuracy of the FV method measured in such a way is compared with the ST method and reported in figure  \ref{fig:KOL}.
While it is known and expected that  accuracy carries some form of dependency with the relaxation time  $\tau$, we observe that both FV and ST methods reach the maximal accuracy (minimal value of the error denoted $E_{\tau}$) around $\tau=0.5$. Such behaviour has been already reported in the works of Holdych \textit{et al.} \cite{Holdych}  and Kruger \textit{et al.} \cite{Kruger}, however ST in that very same case performs better by a factor 10.  Moreover, in general the ST error grows less than the FV one for all $\tau<0.5$.

\subsection {Steady Poiseuille flow}
\begin{figure}
  \centering
           \includegraphics[width=0.5\textwidth]{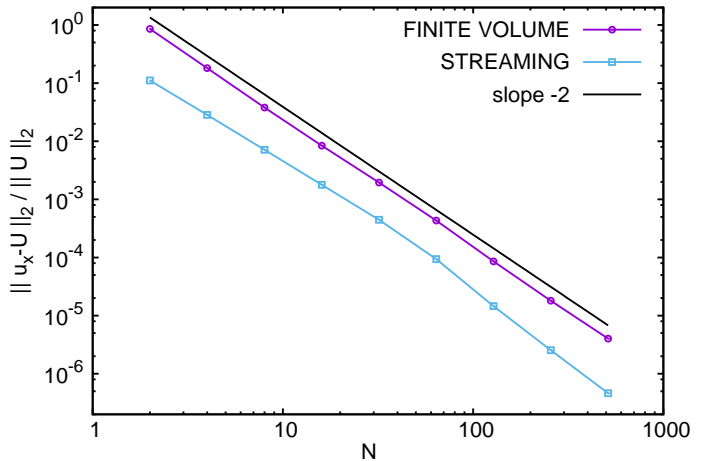}
  \caption{ Relative error on the $Re =10$ velocity Poiseuille flow profile at changing the number of grid points ($N$) and keeping fixed the grid spacing $\Delta x  \equiv L_y/N_y = 1$. Proof that FV has same order of incompressibility accuracy of ST but is less accurate of a factor 8 to 10.}
  \label{fig:POI}
\end{figure}

Our second tests addresses the case of a simple bounded flow in the same spatial domain as above, $[L_x,L_y,L_z]=[1,64,1]$. The  flow is initiated with a parabolic Poiseuille velocity profile $U_x(y) = 4\ U_{max} L_y^{-2}\ y\ (L_y - y)$ corresponding to a Reynolds number $Re=L_y U_{max}/ \nu = 10$. 
A uniform volume forcing along the $x$ direction and no-slip boundary conditions at $y=0$ and $y=L_y$ positions are imposed, while periodicity is implemented along the horizontal direction, $x$. The forcing is implemented via a constant acceleration $a_{x} = \frac{4}{L_{y}^{2}}\ \frac{\nu} {\rho} \ U$ using equation (\ref{eq:F2}).
The simulated flow profile (denoted with $u_x(y)$) keeps the original theoretical shape $U_x(y)$ with tiny adjustments depending on the method.
In order to compare these two functions we use the relative difference $|| u_x - U ||_2\ /\ || U ||_2   $ where 
$||\ldots ||_2$ denotes the $L_2$ norm, which in its discretised form is computed as: 
\begin{equation}
|| f(x) ||_2 = \left( \int_{L} f(x)^2 dx \right)^{1/2}  = \left(\Sigma_{i=1}^{N} f_i^2 \Delta x_i \right)^{1/2} 
\end{equation}
In figure \ref{fig:POI}, we show the $L_2$ relative difference results at varying the  physical domain size in $y$ direction, \textit{i.e.}, changing $L_y$ and  at the same time $N_y$ (or in other words keeping fixed the grid spacing $\Delta x  \equiv L_y/N_y = 1$). Note that in this test we are actually varying the maximal Mach, $Ma=U_{max}/c_s$, number of the flow. Since the LB equation is $\mathcal{O}(Ma^2)$ accurate \cite{Succi2001book}, the figure proves that both the FV and ST methods posses the same level of incompressibility accuracy.
However, we can clearly notice that ST in this conditions is still on average more accurate by a factor $\sim 8 - 10$ as compared to FV, such  a  difference is inherited from the behaviour at $\tau=0.5$ of the viscosity accuracy highlighted in the previous test.

As further step, we address the effect of a stretched spatial grid on the overall accuracy of the Poiseuille flow simulation. 
To this end we implement three types of commonly used 
wall-normal stretched grids.  The $y$-coordinate value of the cell volume centres (or simply nodes) is given by 
$$  y_i =  \frac{ \xi_{i+1} + \xi_{i} }{2}  \qquad  \textrm{with}\  0\leqslant i < N_y$$
where the $\xi_{i}$, the coordinates at the volume boundaries, are defined as
\begin{widetext}
\begin{eqnarray}\label{eq:grids}
 \textrm{Chebychev nodes:} & \xi_{i} & = \frac{L}{2}\left( 1 -   \cos \left[\frac{(i\ -\ 1/2)\pi}{N}\right] \right)  \quad \quad \textrm{where}\ 0\leqslant i \leqslant N\\
\! \! \!  \textrm{hyperbolic tangent:} &   \xi_i & = \frac{L}{2}  \left(1+\frac{1}{s_1} \tanh \left[(\frac{2}{N} \  i -1 ) \  \textrm{atanh}(s_1)\right]\right) \quad \text{$\textrm{where}\  0\leqslant i \leqslant N$}\\
 \textrm{hyperbolic sine:} &  \xi_i &= \begin{cases}
     \left(\frac{L/2}{\sinh(s_2/2)}\right) \sinh \left(\frac{s_2 \  i}{N}\right),                                                                     &\text{$\textrm{if} \ 0\leqslant i\leqslant \frac{N}{2}$}\\
     L - \left(\frac{L/2}{ \sinh(s_2/2)}\right) \sinh \left(\frac{s_2 \ (N-i)}{N}\right),                                                                &\text{$\textrm{if} \ (\frac{N}{2}+1)\leqslant i \leqslant N$}.
 \end{cases}  
\end{eqnarray}
\end{widetext}
$L$ and $N$ denote here respectively the physical domain size and the grid size (the sub-script index $y$ have been dropped for brevity), and $s_1, s_2$  are stretching factors (we have chosen $s_1=0.98$ and $s_2=6.5$).
We then perform the same, $Re =10$, Poiseuille flow simulation with the three above different grid arrangments with the FV method, and for completeness we also include the results obtained on a uniform grid by both the FV and the ST method. In order to have a better understanding on the accuracy of the methods this time we 
change the number of grid points ($N$) while keeping fixed the physical domain size $L=64$. In other words what we vary here is the average grid spacing  $\langle \Delta x \rangle = L/N$.

\begin{figure}[H]
  \centering
           \includegraphics[width=0.5\textwidth]{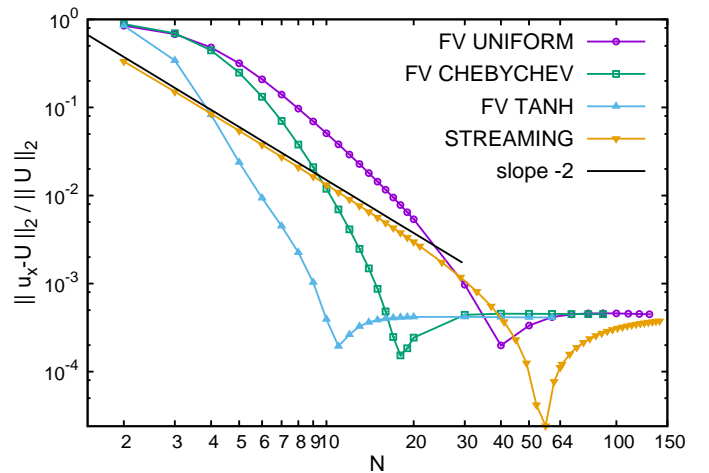}
  \caption{ Relative error on the $Re =10$ velocity Poiseuille flow profile at changing the number of grid points ($N$) and keeping fixed the domain size $L=64$. 
  Data are traced for FV with uniform grid, with Chebychev points and hyperbolic tangent grid refinement, and finally for the ST algorithm with uniform grid.}
  \label{fig:POI2}
\end{figure}

Figure \ref{fig:POI2} reports the results of the described test. Three sources of  inaccuracy lead to the overall error behaviour observed here.  The asymptotic behaviours are respectively dominated by the spatial accuracy error $E_{\Delta x}$ and by the finite Mach correction $E_{Ma}$, while the transition between these two regimes is also affected by the relaxation time error $E_{\tau}$. At small resolutions (small values of $N$ or equivalently, values of $\Delta x>1$ in our numerical experiment) the spatial discretisation error of ST method goes as $E_{\Delta x} \sim \mathcal{O}(\Delta x^2)$ while the one of the uniform-grid FV method behave roughly as $E_{\Delta x} \sim \mathcal{O}(\Delta x^3)$ (due to the QUICK flux computation). In the opposite limit (large $N$, or equivalently $\Delta x < 1$) the compressibility corrections comes into play. This effect which goes as $E_{Ma} \sim \mathcal{O}(Ma^{2})$ has a constant behaviour, both in the ST and FV method, because in this numerical experiment $U$ is kept fixed. This explain the observed plateau in the same figure. At the crossover between the two regimes, around the value $\Delta x \simeq 1$ it also happens that $\Delta t \simeq 2 \tau$  and this corresponds to the range of best accuracy on the viscosity term (minimal $E_{\tau}$) (same as in figure \ref{fig:KOL}).  In conclusion, there exist an optimum value of $N$  linked to the relaxation time error $E_{\tau}$ for which the error is minimum,
this happens both for the FV and ST algorithms, both on uniform and on stretched grids. The ST method (which can only be based on a uniform grid) performs better than the uniform-grid FV implementation  for almost all the values of  $N$, furthermore its absolute accuracy is the highest. However, the situation becomes interesting when the non-uniform grid FV method is employed. There we notice that one can get the same accuracy of the ST algorithm but with a smaller amount of grid points. For instance, in the case of the hyberbolic-tangent grid with $N=11$  one can obtain the same accuracy as the ST algorithm with $N \simeq 46$. This leads to a saving in memory and potentially in computational costs. In conclusion, the reduction in memory occupation at comparable accuracy seems to be the main benefit on can get from employing  the wall-refined FV method rather than the standard ST. \\

\begin{figure}[H]
  \centering
           \includegraphics[width=0.5\textwidth]{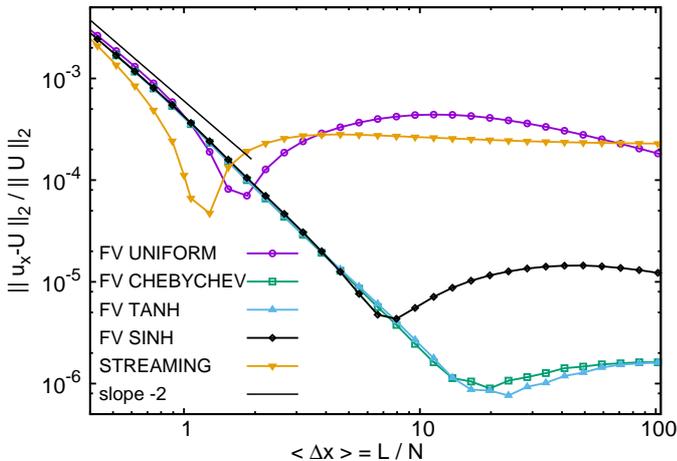}
  \caption{ Relative error on the $Re =10$ velocity Poiseuille flow profile at changing simultaneously the domain size $L$ and the forcing amplitude, but keeping constant the number of grid points $N = 64$. Here the data are plotted versus  the average grid spacing  $\langle \Delta x \rangle = L/N$.  
  Data are traced for FV with uniform grid, with Chebychev points, hyperbolic tangent and hyperbolic sine grid refinement. The corresponding relative error for the ST algorithm with uniform grid is also traced.}
  \label{fig:POI3}
\end{figure}

However, a situation that often occur in the simulation practice is that one wants to use all the available memory of a computer and using an algorithm with the best possible accuracy.  The interesting question is then:  How can we increase the accuracy at comparable memory costs?
Let's imagine one wants to perform again the same Poiseuille simulation at $Re=10$ but wants to reach a higher level of accuracy (with accuracy defined in the sense of $L_2$ norm). One new possibility is to adjust $L$ and $U_{max}$ in a way that the averaged grid spacing  $\langle \Delta x \rangle = L/N$, with $N$ left unchanged,  is the one that offers the best accuracy performance for a given grid. For the above case of the hyberbolic-tangent grid this would be around $\langle \Delta x \rangle = L/N = 64/11 = 5.8$. The result of this novel Poiseuille flow test is shown in figure  \ref{fig:POI3}. We can see that when $N=64$ the best choice is to adopt a grid with $tanh$ or Chebychev spacing and with $\langle \Delta x \rangle$ much larger than unit. The optimum is here  reached when $\langle \Delta x \rangle \simeq 20$, this produce an increase in accuracy of a factor greater than 100 compared to the case of a simulation with the ST algorithm (and this independently of the value of  $\langle \Delta x \rangle$ chosen for the ST method).

\section {Performance evaluation}\label{sec:perf}
\begin{figure}
  \centering
              \includegraphics[width=0.5\textwidth]{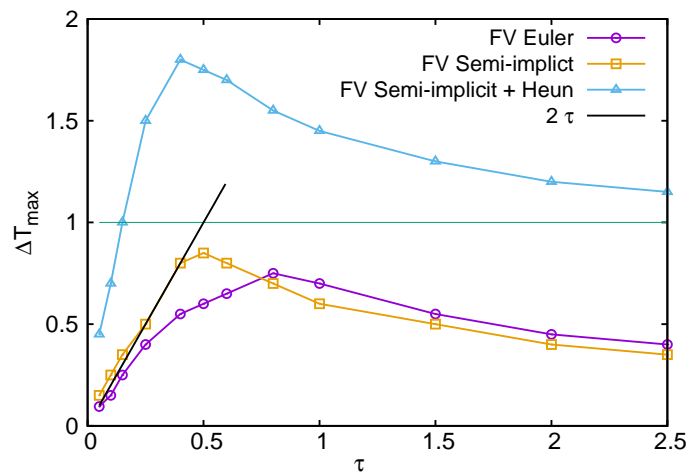}
  \caption{Maximum allowed time step in the decaying Kolmogorov flow by using eq. (\ref{eq:euler}) (FV Euler),  eq. (\ref{eq:semi})  (FV Semi-implicit) and eq.(\ref{eq:semi-heun}) (FV Semi-implicit + Heun). The very same result is obtained in the steady Poiseuille flow at $Re=10$. The horizontal dashed line represents the standard choice of the time-step for the ST  implementation, \textit{i.e.}, $\Delta t =1$ independently of $\tau$. }
  \label{fig:KOL3}
\end{figure}

From a computational point of view the FV algorithm has more operations per time step than the ST algorithm. This comes from the fact that while the streaming process can be implemented simply as a shift in memory the computation of the flux term in FV involves many arithmetics operations.
According to our measurement  the present FV algorithm is about 8-10 times computationally more expensive than ST algorithm per time step.   
However, as discussed in section \ref{sec:time}, differently from the ST algorithm, in the FV the time-step $\Delta t$  is a function of $\tau$. 
The functional relation linking the maximum time-step to $\tau$ for the proposed time-discretizations can be measured and it is reported in figure \ref{fig:KOL3}.
We observe that the method based on (\ref{eq:semi-heun}), semi-implicit integration in time of the collision term plus a trapezoidal correction for the advection is superior to the others.
In particular, for this method $\Delta t_{max} > 1$ for $\tau > 1/8$, that is to say that the time-step can be larger than the one used in the ST method (which is bounded to the value 1 for $\Delta x = 1$). The most advantageous case occurs for $\tau \sim1/2$ - which as we have shown above  is also  the best condition for accuracy - in that case $\Delta t_{max} \sim 1.7$. This reduces the ratio of the  computational  cost FV/ST to a factor 5-6. We note that all this reasoning did not take into account the effect of non-uniform grids. As we have seen for the simple Poiseuille flow this bring further saving in terms of computational costs as compared to the ST method.  

Finally we shall mention that memory occupation is also part of the performance of an algorithm: According to our estimate FV in the present formulation needs  twice more memory allocation as compared to the ST.

\section{Benchmark in a complex flow: high Rayleigh number thermal convection}\label{sec:rb}
Several LB Finite-Volume methods proposed in the past have been tested just on laminar flows as proof of principle of the proposed algorithms.   Few exceptions exist in the literature in which the FV method have been  benchmarked on much more complex, three-dimensional, developed turbulent flows.  One of such exception is the model proposed by Amati \textit{et al.} \cite{Amati}, which was probed in a three-dimensional plane turbulent channel flow. In such case however, the grid wall refinement was based on a very simple structure of halved-grid spacing near the walls and the accuracy of the method turned out to be not satisfactory (the computed mean-velocity profile could not properly reproduce the log-law of the wall). 

In this section the proposed Lattice Boltzmann FV algorithm is tested to simulate a complex three-dimensional statistically steady turbulent flow.  Our choice is here for the well-studied flow in the Rayleigh-B\'enard (RB) cell, the prototype of thermal convection driven system \cite{Goetling}. The RB set-up considered in this study deals with a cubic domain (of height $H$ and equal lateral sizes $L$); it has periodic BC on the lateral walls, while on the horizontal walls no-slip and isothermal conditions are imposed.  In this system, the fluid is heated from below and as such (when the heating is large enough and a small perturbation is introduced in the system) an instability arises and brings the system into convective condition. The dimensionless control parameters are the Rayleigh $(Ra)$ and Prandtl ($Pr$) numbers and aspect-ratio $Ar=L/H$ \cite{Goetling}. 
 
For the LB simulation we use a double population approach \cite{ShanRB}.  This means that beside equation (\ref{eq:LB})  we integrate an analogous equation for the distributions $g_{\alpha}$:
\begin{equation}\label{eq:LBg}
 \frac{\partial g_{\alpha}}{\partial t}+ {\bm c}_{\alpha} \cdot {\bm \nabla} g_{\alpha} = \frac{1}{\tau_g} (g_{\alpha}^{eq} - g_{\alpha})  \qquad   \alpha = 0,\ldots , N_{pop}
\end{equation}
with equilibrium function $g^{eq}_{\alpha} = (T/\rho)\ f^{eq}_{\alpha}$ where the macroscopic temperature is computed as $T = \Sigma_{\alpha} g_{\alpha}$ and  the thermal diffusivity corresponds to $\kappa = \tau_g\ c_s^2 $.
Furthermore, in the equation for $f_{\alpha}$ the forcing term $F_{\alpha}$ is assigned in order to model the buoyancy force as represented in the Boussinesq approximation. In physical space the added buoyant acceleration  has the form
${\bm a}  = - \beta (T-T_0) {\bm g}$ where $\beta$ is the volume thermal expansion coefficient and $T_0$ is a reference temperature taken here as the mean temperature between the ones  at the top and bottom plates. 

\begin{figure}[h]
  \centering             
               \includegraphics[width=0.45\textwidth]{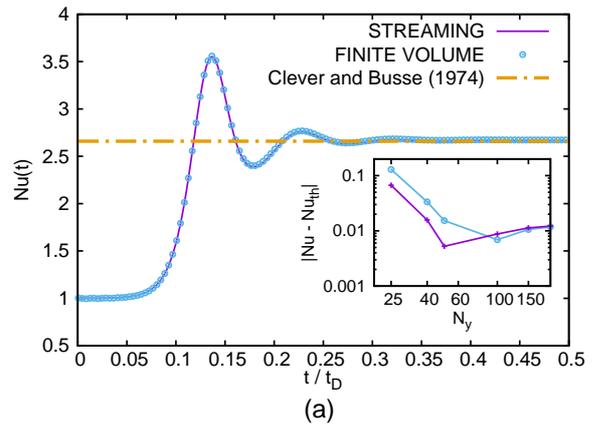}
              \includegraphics[width=0.5\textwidth]{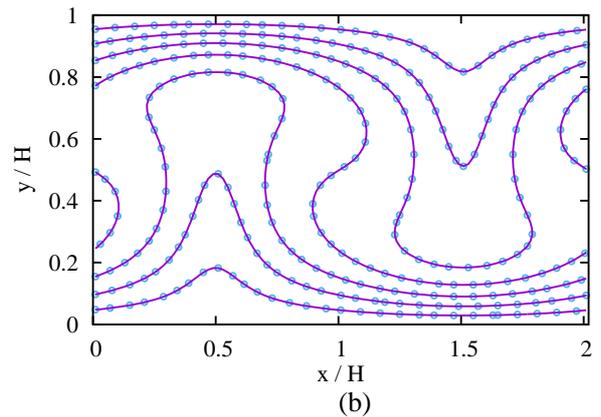}
  \caption{Comparison of streaming (ST) and finite-volume(FV) LB algorithms in a simulation of the Rayleigh-B\'enard system in steady convective state. The system is two-dimensional, and characterised by 
  the control parameters value $Ra = 10^4$, $Pr = 1$ and $Ar=2.02$. (a) Temporal dynamics of dimensionless global heat flux (Nusselt number $Nu(t)$) as a function of time, in dissipative time units $t_D=H^2/\kappa$. The $Nu$ steady state value is compared to a value linearly interpolated from Clever and Busse calculations \cite{Clever}. In the inset, the grid convergence study displaying the absolute error of the measured Nusselt number $(Nu)$ with respect to Clever and Busse $(Nu_{th})$ $vs.$ number of grid points in the y-direction of a 2D grid. (b) Comparison of temperature isolines in the asymptotic steady state. Levels are taken at values $T_{n} = T_0 \pm n\ \Delta T/8$, with $n=0,1,2,3$.}
\label{fig:RBlow}
\end{figure}

\begin{figure}[hbt]
  \centering
              \includegraphics[width=0.47\textwidth]{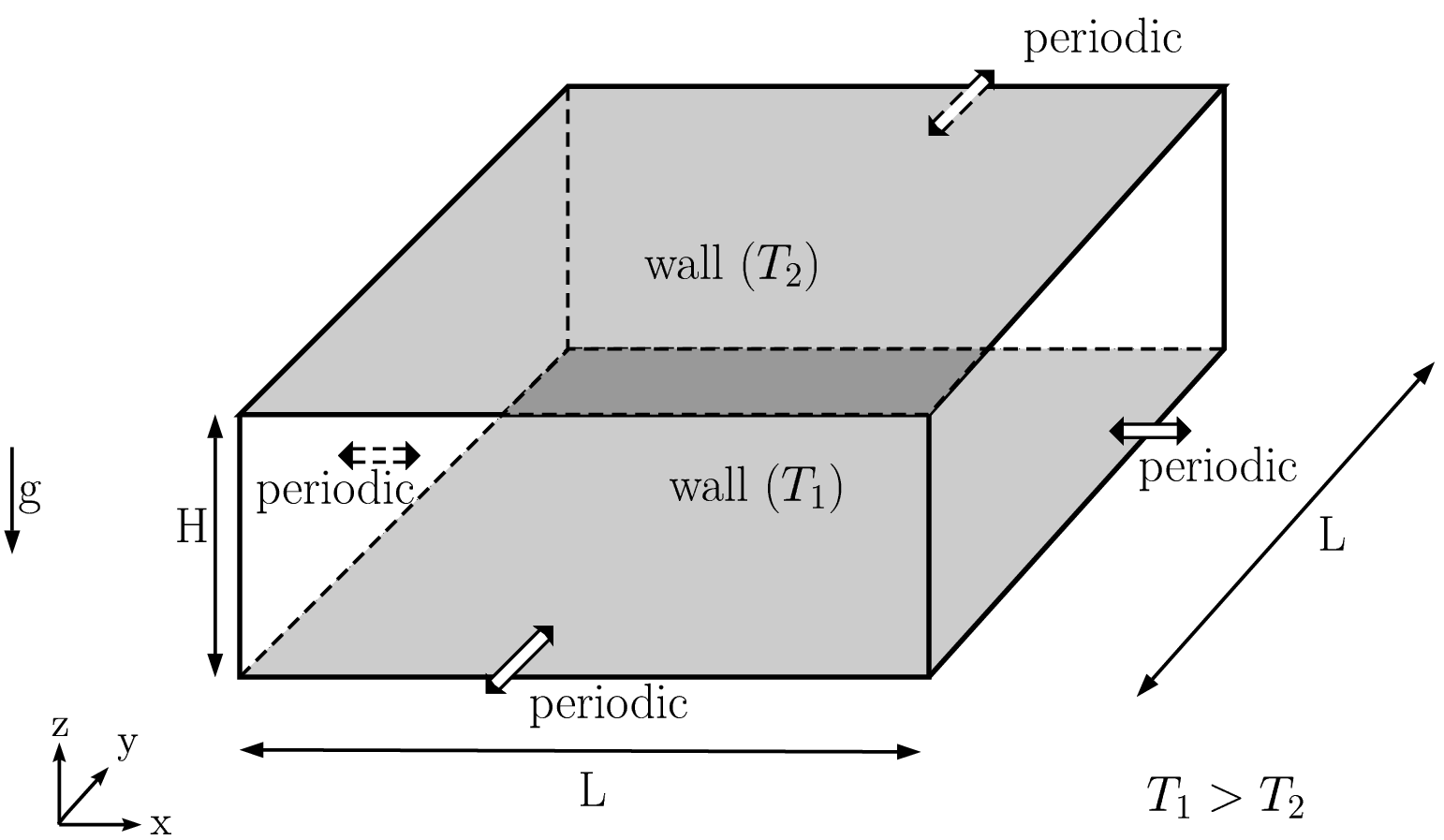}
  \caption{Cartoon of the three-dimensional Rayleigh-B\'enard system}
  \label{fig:RB}
\end{figure}

In order to validate the double population approach also for the FV method, we first address a rather elementary simulation in steady convective laminar conditions, adapting it from a test case already conducted for the ST algorithm in Ref. \cite{ShanRB}. The system is two-dimensional (2D) with control parameters  fixed at $Ra=10^4$, $Pr=1$ and $Ar=2.02$. 
The fluid is initially at rest (${\bm u} = 0$), while the temperature field is initialised by a linear conductive profile, $T_c(z) = -\Delta T (z/H + 1/2)$, plus a small perturbation (of order $O(10^{-2}) \Delta T$) breaking the left right symmetry. 
Given the weak, but not negligible, compressibility of the simulated flow the initial density stratification due to gravity should be also taken into account. This avoids the generation of pressure waves at the startup of the simulation. We do it via the barometric equation, this leads to $\rho(z) = \rho_0 \exp{( -c_s^{-2}\beta g  \int_0^z T_c(z') dz')}$, where $\rho_0$ is a reference density value taken at temperature $T_0$. 
Note that in a 2D system, in order not to suppress the linear hydrodynamic instability, the cell aspect ratio ($Ar$) must be slightly larger than $2 \pi / k_c$  (where $k_c = 3.117$ is the wave vector of the most unstable linear mode) \cite{ShanRB}. Indeed, when $Ar=2.02$ the initial perturbation produce an immediate kinetic energy growth and a steady convective flow pattern establishes. The dimensionless heat flux (or Nusselt number $Nu$) goes from the conductive unit value up to around $Nu \simeq 2.66$, see \cite{Clever}.
In figure \ref{fig:RBlow} we report the results of simulations conducted with the two LB algorithms. We can observe (Fig. \ref{fig:RBlow}a) that the temporal dynamics of the dimensionless global heat flux, \textit{i.e.} the Nusselt number $Nu(t)$, is identical for the two simulations, furthermore they both agree with the analytical asymptotic value given by Clever and Busse \cite{Clever}. A grid convergence study, performed by increasing the number of grid points of the same factor in each cartesian direction, shows that the absolute error on Nusselt number as compared to the Clever and Busse value reaches the second decimal digit (inset of Fig. \ref{fig:RBlow}a). The isocontours lines for the temperature field (Fig. \ref{fig:RBlow}b)  further display the excellent agreement between the two LB algorithms. The test exhibits not only the good quality of the present FV method but also its consistency with the standard ST method also for transient (\textit{i.e.} time-dependent) dynamics. 

 \begin{figure}[h]
  \centering
              \includegraphics[width=0.5\textwidth]{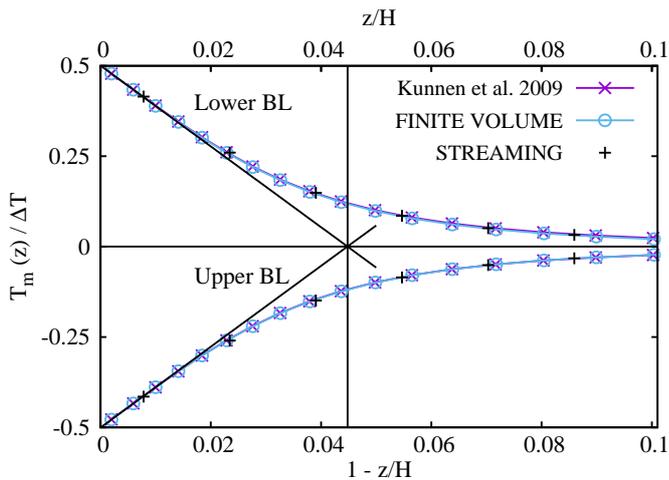}\\
  \caption{The mean temperature profile $T_m(z)$ - averaged over time and horizontal planes - as a function of the height $z$ in the cell.  To better appreciate the agreement between different simulation methods we show here a close-up view of the profiles in lower/upper $10\%$ of the cell.}
  \label{fig:cfrT}
\end{figure}

\begin{figure}[h]
  \centering             
              \includegraphics[width=0.5\textwidth]{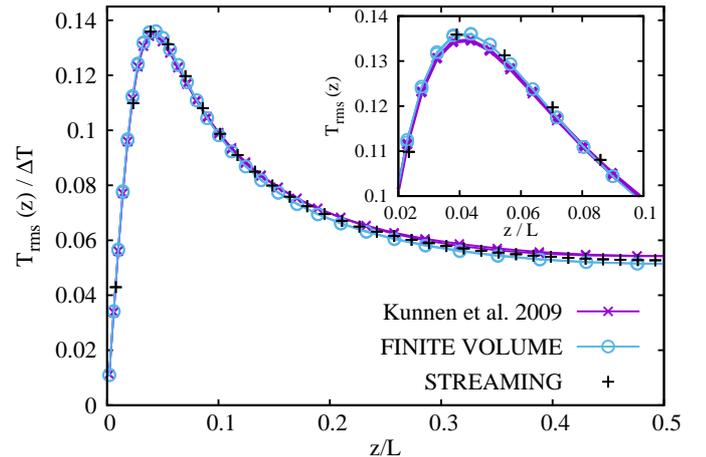}
  \caption{Root-mean-square temperature profiles $T_{rms}$, averaged over time and horizontal planes, as a function of the cell height $z$ up to the cell center height. In the inset, a zoomed in view of the lower $10\%$ of the cell.}
  \label{fig:cfrT2}
\end{figure}

We then move forward to a more complex case. In particular, we compare our results with the ones obtained by Kunnen \textit{et al.} \cite{Kunnen}  for a three-dimensional (3D) simulation of a RB system (Fig. \ref{fig:RB}) characterised by: $Ra = 2.5 \cdot 10^6$, $Pr = 1$ and $Ar=2$ (see also \cite{Lavezzo}). In this condition the 3D system dynamics is already highly chaotic (or moderately turbulent).
In \cite{Kunnen} the authors employed a direct numerical simulation  based on a staggered finite-difference discretisation of the Navier-Stokes - Boussinesq equation system. The grid they adopted has size $(N_x,N_y,N_z) = (128,128,64)$, it is uniform in the horizontal directions and has a $sinh$-type refinement (the same as in (\ref{eq:grids})) in the vertical direction.   Our benchmark is as follow, we perform two series of simulations, one with the ST method and the other with the FV approach, the dimensionless parameters for the two cases are the same as the ones of Kunnen \textit{et al.},  as well as the number of grid points per direction. However, while the ST uses a uniform grid in the FV case we use exactly the same grid as the one adopted in the finite-difference simulation \cite{Kunnen}.  The table \ref{tab:rb} reports the numerical values of the parameters adopted for the two LB simulations.  Note that the large scale velocity $U$ which is roughly proportional to the so called free-fall velocity, \textit{i.e.} $U \sim \sqrt{ \beta g \Delta T H}$  is the same in both simulations.  It is a good practice in LB simulations to always keep control of the large-scale velocity in order to prevent it to take too large values: it is worth reminding that in order to reproduce the incompressible fluid-dynamics the condition $U \ll 1$ is required (a commonly accepted rule of thumb in LB practice is $U \simeq 0.1$). In order to reach a good convergence of the statistical observables in the system the RB simulations are carried on for a total time ($t_{tot}$) which spans over several large eddy turnover times ($\mathcal{T}$). We estimate that $t_{tot} \simeq 12\ \mathcal{T}$ for both FV and ST simulations, with  $\mathcal{T}$  computed from the zero-crossing time value of the autocorrelation function of  the total kinetic energy. 

\begin{table}
\begin{center}
\begin{tabular}{cccccccccc}
\hline
      &$H$ & $L$ &$\Delta t$ & $\tau$       & $\tau_{g}$  & $\Delta T$&  $\beta$ & g  & $t_{tot}$\\
\hline
FV  & 640 &   1280 &    4  &            0.5 & 0.5             & 2               & $1.325\cdot 10^{-4}$     & 1 & $1.28 \cdot 10^6$ \\
\hline
ST  & 64 &    128 & 1 & $0.05$ & $0.05$ & 2              & $1.325\cdot 10^{-3}$      & 1 &  $2.48 \cdot 10^5$ \\
\hline
\end{tabular}
\end{center}
\caption{Parameter values used for the RB simulations at $Ra=2.5 \cdot 10^{6}$, $Pr=1$ and $Ar=2$: height ($H$) and width ($L$) of the cell, time step amplitude $\Delta t$, 
fluid ($\tau$) and temperature ($\tau_g$) relaxation times, temperature gap across the cell $\Delta T$, thermal expansion coefficient value $\beta$ and intensity of gravity $g$. $t_{tot}$ is the total simulation time in numerical units.} \label{tab:rb}
\end{table}

In the figures \ref{fig:cfrT} and  \ref{fig:cfrT2} we show a comparison of the vertical mean temperature profile ($T_m$) (averaged over horizontal planes and time) and  of the vertical root-mean-square temperature ($T_{rms}$) profile, which are defined as follows: 
\begin{widetext}
\begin{eqnarray}
T_m(z) &=&  \frac{1}{t_{tot} \ L^2} \int_0^{t_{tot}} \int_0^{L} \int_0^{L}   T(x,y,z,t)\ dx \  dy \ dt \\
T_{rms}(z) &=&  \left( \frac{1}{t_{tot} L^2}\int_0^{t_{tot}}  \int_0^{L}  \int_0^{L}   (T(x,y,z,t) - T_m(z) )^2 \ dx \  dy\ dt \right)^{1/2}.
\end{eqnarray}
\end{widetext}

\begin{figure}[!hb]
  \centering             
              \includegraphics[width=0.5\textwidth]{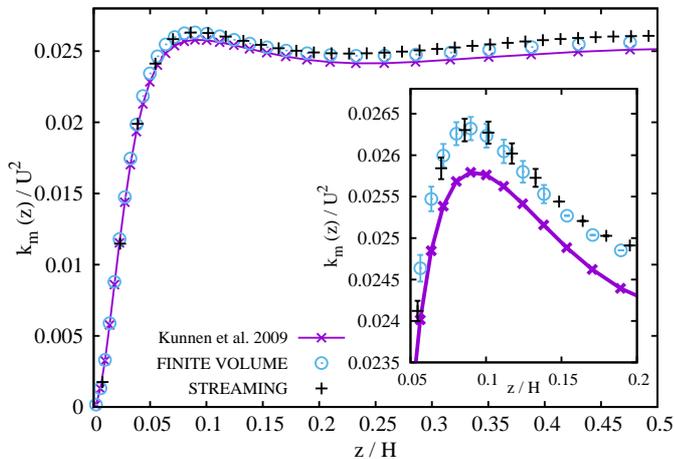}
  \caption{Mean turbulent kinetic energy $k_m(z)$, averaged over time and horizontal planes, as a function of the cell height $z$, 
  and close-up view around the BL peak value (inset).}
  \label{fig:cfrE}
\end{figure}

\begin{figure}[h]
  \centering             
               \includegraphics[width=0.5\textwidth]{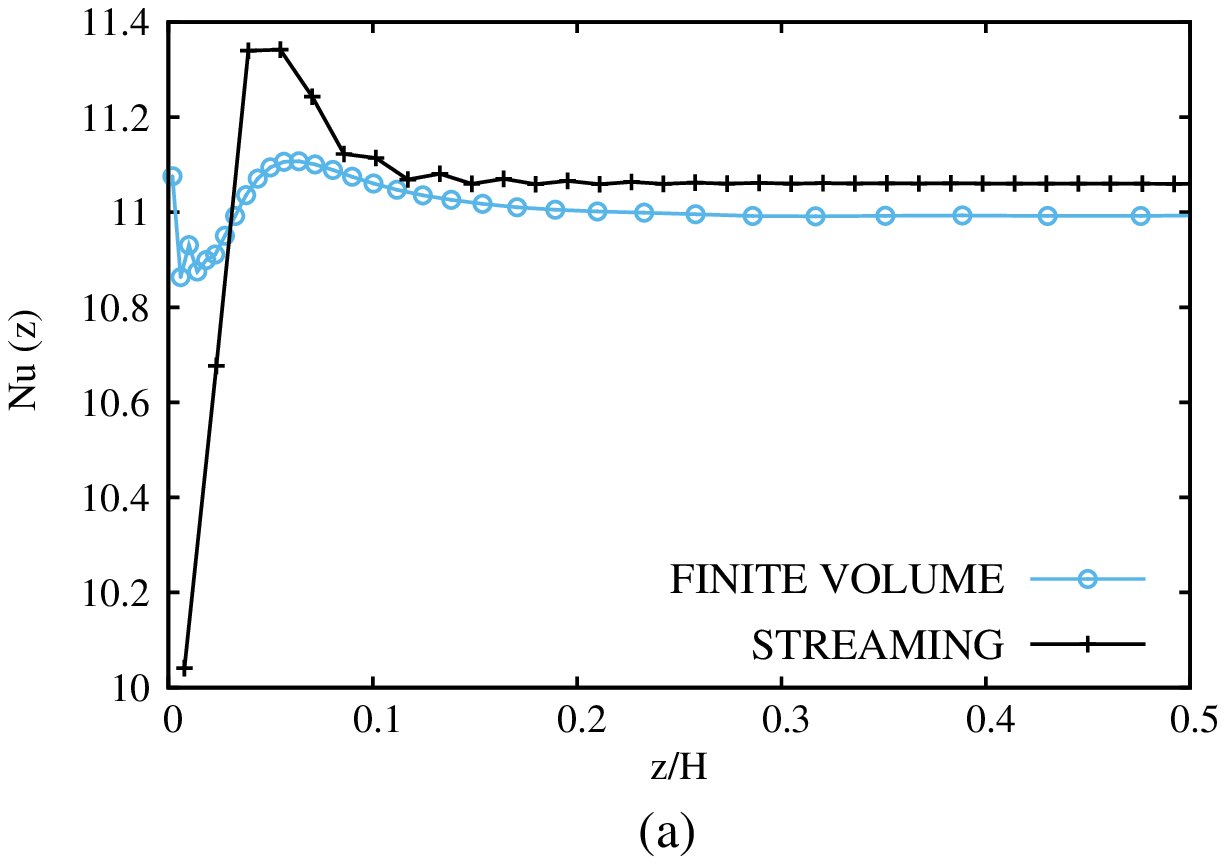}
              \includegraphics[width=0.5\textwidth]{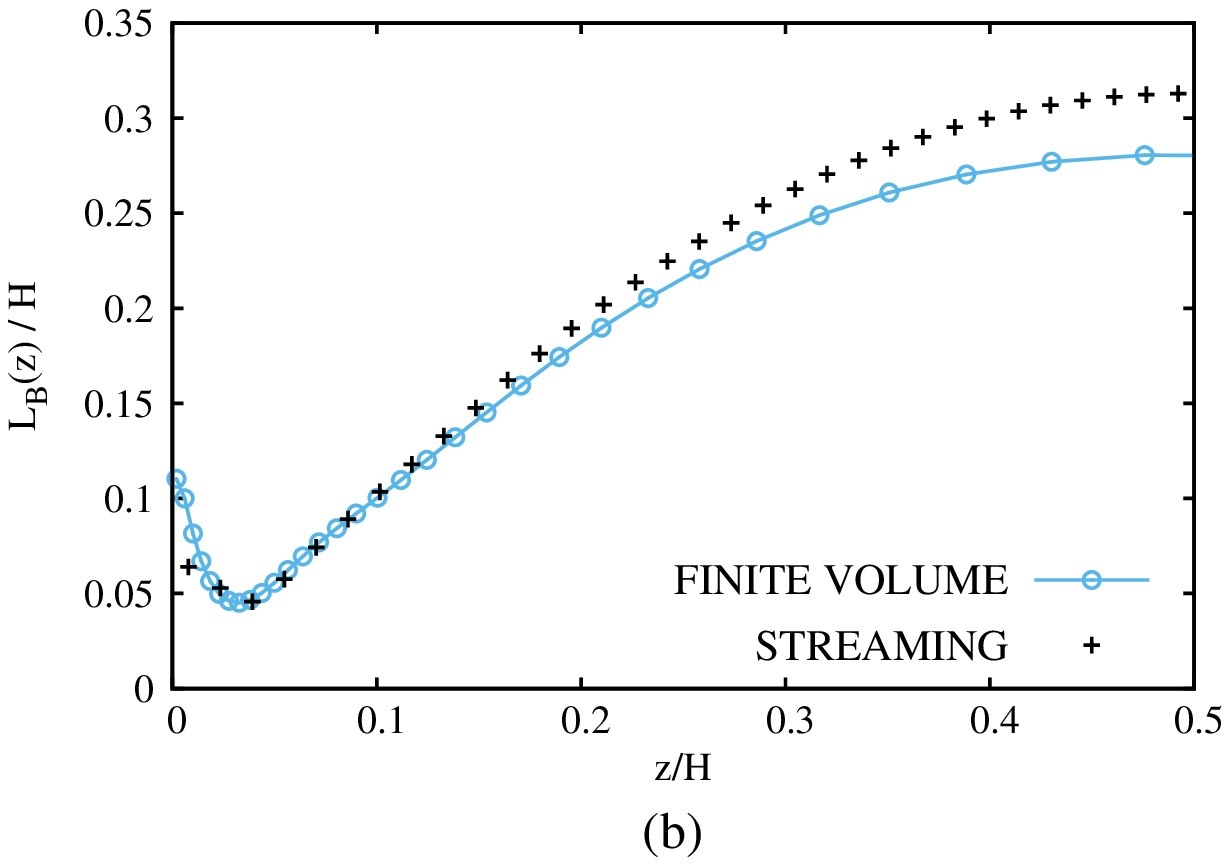}
  \caption{(a) The time average Nusselt number $Nu(z)$ as a function of the height in the cell, up to half cell. Note that in a ideal RB system this quantity should be constant, however in numerics, often due to the effect of BC implementations, small fluctuations are observed through the cell. It is here evident  the higher quality of the FV method.
  (b) Bolgiano length, $L_B(z)$, averaged over time and horizontal planes, as a function of the cell height $z$. Note the discrepancies both at the wall and in the cell bulk.}
  \label{fig:cfrLB}
\end{figure}

We find good agreement among all the three types of simulations. Furthermore, we observe that when the thickness of the boundary layer $\lambda_T$ is defined by the so called slope definition, $\lambda_T \equiv \Delta T (2\  \partial_z T_m(z) |_{z=0} )^{-1}$ (see figure \ref{fig:cfrT}) both the Kunnen \textit{et al.} data and the FV ones have about 10 points in the thermal boundary layer (BL), while the ST despite its remarkable agreement with the other methods, has only 3 points in the BL.  Small systematic differences can be seen on the vertical profile of  the mean (turbulent) kinetic energy, $k_m (z) =   t_{tot}^{-1}\ L^{-2} \int_0^{t_{tot}}  \int_0^{L} \int_0^{L}   \frac{1}{2}  \bm{u}^2 \ dx \  dy \ $ , reported in figure \ref{fig:cfrE}.  $k_m (z)$ has a slower rate of convergence than the temperature variance, this is the reason why small residual statistical discrepancies remain present here despite of the large number of turnover times of the simulation.  In order to appreciate more sensible differences between the FV and ST simulation one has to address either observables involving temperature-velocity correlation or small-scale quantities, which are more sensitive to the spatial resolution of the mesh, particularly close to the walls. For this reason in figure \ref{fig:cfrLB} we compare the time averaged quota-dependent Nusselt number $Nu(z) = \kappa \partial_z \langle T \rangle+ \langle u_z T\rangle  / ( \kappa \Delta / H)$ (where for short $\langle \ldots \rangle$  denotes time and space horizontal averages) and the so called Bolgiano length $L_B(z) = (\beta \ g)^{-3/2} \langle \epsilon_u\rangle ^{5/4} \langle \epsilon_T\rangle ^{-3/4}$ (where $\epsilon_u$ and $\epsilon_T$ are respectively the velocity and temperature dissipation rates) \cite{Calza}.  As far as Nusselt number is concerned, despite a very close mean value, we see important differences at the wall. This is due to the combined effect of the boundary conditions and the gradient computations in post-processing the data. The temperature gradient computations involved a second order central finite difference scheme at the bulk nodes and a second order forward/backward finite difference scheme at the boundary nodes. These schemes have been applied to both FV and ST algorithms. The FV method exhibit wall oscillations which are a factor 10 smaller than the ones seen for the ST method, making more reliable the  total heat flux estimate.  Furthermore, in the $L_B(z)$ measure we observe near a $50 \%$ discrepancy at the wall and a smaller but non negligible difference in the bulk of the cell. Clearly a wall-clustered grid is needed to resolve observables built on sharp temperature and velocity gradients.  

We now would like to address the matter of determining which LB method is computationally more convenient.  
The choice to have the same $U$ in the FV and ST simulations has an implication on the determination of the large-eddy turnover time  and therefore the total number of time steps needed to perform a simulation of equivalent physical time-span. 
The reasoning is as follow: the large turnover time  goes as  $\mathcal{T}  \sim  H/U$ therefore on the total number of time-steps $M$ for a simulation that should span a time $\mathcal{T}$ scale as  $M \sim H/\Delta t$. It follows that the FV simulation will need in this case a number of time steps larger by a factor $10/4$ as compared to the ST one (see table \ref{tab:rb}). 
Since the FV is more expensive than ST by a factor $8-10$ per time step, we get that the added computational cost of the FV method is up to $ \simeq 25$ larger than the ST method.
However, such an increase in computational cost shall be properly weighted by the enhancement in the spatial resolution due to the wall stretched grid.  
An univocal guideline is not available in this context. A commonly employed criterion in the numerics of bounded flows is to count the number of grid nodes in the BL (another, although less restrictive, rule would be to take into account the distance of the first collocation point from the wall). Here, if we adopt such a criterion the ratio is in favour of the FV method over ST by a factor $10/3$, that means that we need approximately 3 - 4 more nodes in the ST simulation.
However if we want also keep the same aspect-ratio of the simulation domain, and since ST is bounded to cubic grids, such an increase of the resolution shall be applied to every cartesian direction,  which makes the FV grid advantage greater of a factor of $(10/3)^3 \simeq  37 $. In conclusion, in a ST RB simulation we need 37 times more computation nodes to perform a simulation with comparable resolution of the FV method.  
By combining the above estimates, we see that a simulation of same physical time-span and same boundary layer resolution, is about $1- 25/37 \sim 33 \%$ less expensive for the FV method than the ST one. 
In summary, even if the cost per unit physical time in a FV simulation is higher than ST, when a criterion for the minimal spatial resolution (particularly near walls or obstacles and in a three dimensional geometry) is chosen, the FV method becomes advantageous.

\begin{figure}[h]
  \centering             
               \includegraphics[width=0.5\textwidth]{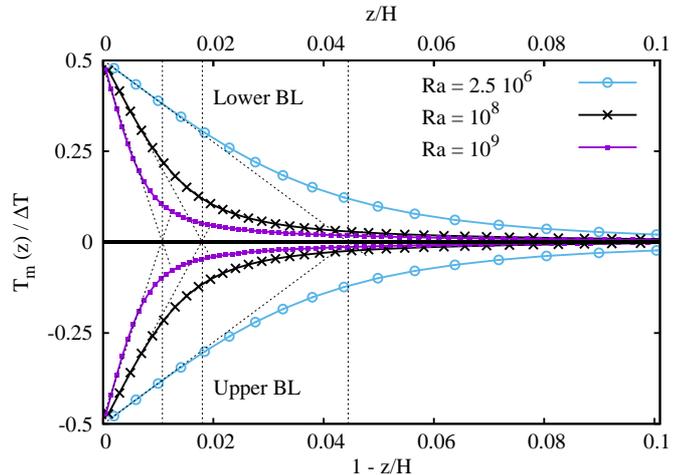}
  \caption{Mean vertical temperature profile, $T_m(z)$, close to the upper and lower plates in the Rayleigh-B\'enard system at  $Ra=2.5\cdot 10^6$ ($\circ$) , $Ra=10^8$ ($\times$) and $Ra=10^9$ ($\square$).
  The Prandl number is $Pr=1$ and aspect ratio $Ar=2$.  The thickness of slope boundary layers is indicated by the dotted lines.
  Results were obtained by the FV algorithm at resolution $64 \times 128^2$, $128^3$ and $256^3$ respectively.}
  \label{fig:ra-high}
\end{figure}

Finally, we have performed RB simulations at increasingly higher $Ra$ numbers $(Ra=10^8, 10^{9})$.
All this simulations have around 10  grid points in the thermal BL as shown in the figure \ref{fig:ra-high}.  
No numerical instabilities were noticeable  as $Ra$ was increased, demonstrating that the FV algorithm can deal with turbulence at highly Rayleigh number conditions.

\section{Conclusions}\label{sec:conc}
In this paper we have presented a new finite volume algorithm for the Lattice Boltzmann equation. The new method has been validated through a systematic comparison with the standard streaming LB approach, by means of test case simulations in laminar as well as in unsteady and turbulent flows with heat transfer.  The tests have shown that the FV has the same order of incompressibility accuracy as the ST algorithm and an improved spatial accuracy (third compared to second order), it has however a much more elevated computational costs that we estimate to be around 8-10 times per time-step. One notable advantage of the FV method is the possibility to adopt stretched rectilinear grids, which makes it suitable for the simulation of turbulent bounded flows. Taking this into account, \textit{i.e.} taking into consideration (for instance) the minimal number of collocation points required in a boundary layer for a proper simulation, the FV algorithm surpasses the ST method.

A number of improvement can still be made on the proposed algorithm. First, the boundary conditions can be improved.
 We have noticed that in strongly sheared flows, such as  channel flow turbulence, some spurious oscillations at the boundaries can destabilise the simulations. 
 Second, in stretched cartesian grids the number of interpolant coefficients to be stored is considerably more limited than for other cases of structured grids. In the former case the treatment of advection can be further optimised compared to the one used in the present study, allowing for some extra saving in computational costs. 
 
The FV discretisation proposed in this work builds on the standard streaming algorithm, as a consequence the two algorithms do not differ much. 
The  major difference is of course the way in which the advection is computed. However, their strong resemblance may be useful in the development of simulation codes that combine the two methods. Therefore, one possible development of the present study is to set-up simulations that use the more efficient ST method in flow domain regions where a fine grid is needed, and the FV method in regions where a more coarse grid will suffice (among others, a typical application could be the simulation of atmospheric boundary layer with ST at ground level and FV on the upper residual layer). It is a prospect that we plan to explore in a future work.

\section*{Acknowledgments}\label{sec:ack}
This work has been supported by a collaboration between academic and industrial partners promoted by the organization  INNOCOLD ({\tt http://www.innocold.org}) and in part by the French National Agency for Research (ANR) by the grant (SEAS: ANR-13-JS09-0010). Rudie Kunnen is kindly acknowledged for sharing the RB simulation results. E.C. thanks Federico Toschi for useful discussions.

\section*{Appendix A.  Mixed semi-implicit Finite Volume LB scheme}
In the following we briefly show how to derive the discretised form (\ref{eq:semi}) from the finite volume LB equation (\ref{eq:fv}). 
First the discretisation in time is applied by adopting a mixed approach: while for the collision and  forcing terms a semi-implicit method is used, for the advection a simple explicit Euler is implemented. This leads to the form:
\begin{widetext}
\begin{equation}
 f_{\alpha}^{(t+\Delta t)}  = f_{\alpha}^{(t)}- \Delta t  \frac{S_j}{V} {\bm c}_{\alpha} \cdot  {\bm{n} }_j  \left[ f_{\alpha}^{(t)} \right]_j + \frac{\Delta t }{2} \left( \frac{1}{\tau} ( f_{\alpha}^{eq\, (t)} - f_{\alpha}^{(t)}) 
+ F_{\alpha}^{(t)}  +  \frac{1}{\tau} ( f_{\alpha}^{eq\, (t+\Delta t)} - f_{\alpha}^{(t+\Delta t)}) +  F_{\alpha}^{(t+\Delta t)}   \right)
\end{equation}
\end{widetext}
Note that by bringing on the left-hand-side all the terms to be evaluated at $(t+\Delta t)$:
\begin{widetext}
\begin{equation}
 f_{\alpha}^{(t+\Delta t)} - \frac{\Delta t }{2\tau} ( f_{\alpha}^{eq\, (t+\Delta t)} - f_{\alpha}^{(t+\Delta t)} +  \tau F_{\alpha}^{(t+\Delta t)} ) = 
f_{\alpha}^{(t)}- \Delta t  \frac{S_j}{V} {\bm c}_{\alpha} \cdot  {\bm{n} }_j  \left[ f_{\alpha}^{(t)} \right]_j + \frac{\Delta t }{2 \tau}  ( f_{\alpha}^{eq\, (t)} - f_{\alpha}^{(t)}
+\tau  F_{\alpha}^{(t)} )
\end{equation}
\end{widetext}
One now introduces the redefined distribution function $\tilde{f}_{\alpha} =f_{\alpha} - \frac{\Delta t}{2\tau} ( f_{\alpha}^{eq} - f_{\alpha} + \tau F_{\alpha})$. 
It shall be noted that, at equilibrium condition, from the above definition we get  $\tilde{f}_{\alpha}^{eq} =f_{\alpha}^{eq} - \frac{\Delta t}{2}  F_{\alpha}$, and therefore the relation can be easily inverted leading to
\begin{equation}
 f_{\alpha} = \tilde{f}_{\alpha} + \frac{\Delta t}{2 \tau + \Delta t}( \tilde{f}_{\alpha}^{eq} - \tilde{f}_{\alpha} ) + \frac{\Delta t}{2} F_{\alpha}
\end{equation}
At this point the new relaxation time is introduced  $\tilde{\tau} = \tau + \frac{\Delta t}{2}$, and the equation can be written as
\begin{widetext}
\begin{equation} 
\tilde{f}_{\alpha}^{(t+\Delta t)} = \tilde{f}_{\alpha} - \Delta t  \frac{S_j}{V} {\bm c}_{\alpha} \cdot  {\bm{n} }_j  \left[ \tilde{f}_{\alpha} +\frac{\Delta t}{2 \tilde{\tau}}( \tilde{f}_{\alpha}^{eq} - \tilde{f}_{\alpha} ) + \frac{\Delta t}{2} F_{\alpha} \right]_j + \frac{\Delta t}{\tilde{\tau}} (\tilde{f}_{\alpha}^{eq} - \tilde{f}_{\alpha}) + \Delta t F_{\alpha}
\end{equation}
\end{widetext}
The above equation coincide with (\ref{eq:semi}), once the correction factor $\left( 1 - \frac{\Delta t}{ 2 \tilde{\tau}}\right)$ is applied to the force intensity $F_{\alpha}$ \cite{Guo}.

\bibliography{citations}

\end{document}